# White Paper

on the case for

# EXOGEOSCIENCE

and its role in

## CHARACTERIZING EXOPLANET HABITABILITY AND THE DETECTABILITY OF LIFE

Cayman T. Unterborn ASU | cayman.unterborn@asu.edu        Paul K. Byrne NCSU | paul.byrne@ncsu.edu

Ariel D. Anbar ASU · Giada Arney NASA GSFC · David Brain CU BOULDER · Steven J. Desch ASU · Bradford J. Foley PSU ·

Martha S. Gilmore WESLEYAN · Hilairy E. Hartnett ASU · Wade G. Henning NASA GSFC · Marc M. Hirschmann UMN ·

Noam R. Izenberg JHU APL · Stephen R. Kane UCR · Edwin S. Kite U CHICAGO · Laura Kreidberg MPIA ·

Kanani K. M. Lee LLNL · Timothy W. Lyons UCR · Stephanie L. Olson PURDUE · Wendy R. Panero OSU ·

Noah J. Planavsky YALE · Christopher T. Reinhard GEORGIA TECH · Joseph P. Renaud NASA GSFC ·

Laura K. Schaefer STANFORD · Edward W. Schwieterman UCR · Linda E. Sohl NASA GISS ·

Elizabeth J. Tasker ISAS JAXA · Michael J. Way NASA GISS

ENDORSERS  VARDAN ADIBEKYAN | CAITLIN AHRENS | HARRISON ALLEN-SUTTER | BRENDAN ANZURES | DANIEL APAI | SARAH ARVESON | MARIA BANKS | RORY BARNES | AMY BARR | DAN BOWER | GRAYSON BOYER | KARALEE BRUGMAN | PETER BUSECK | KEVIN CANNON | LUDMILA CARONE | RICHARD CARTWRIGHT | DAVID CATLING | JULIAN CHESNUTT | ADITYA CHOPRA | WILLAIM COCHRAN | CADY COLEMAN | PAUL DALBA | WILLIAM DANCHI | JIE DENG | CHUANFEI DONG | JEREMY DRAKE | STEPHEN ELARDO | AL EMRAN | NÉSTOR ESPINOZA | JACK FARMER | THOMAS FAUCHEZ | RACHEL FERNANDES | RODOLFO GARCIA | DAWN GELINO | MARINA GEMMA | MARK GHIORSO | KAUSTUBH HAKIM | SEBASTIEN HAMEL | HEIDI HAMMEL | JENNIFER HANLEY | CHESTER 'SONNY' HARMAN | JOSEPH HARRINGTON | ASHLEY HERBST | NATALIE HINKEL | TIMOTHY HOLT | CHRISTINE HOUSER | DEVANSHU JHA | JOHANNA TESKE | BENJAMIN JOHNSON | BAPTISTE JOURNAUX | BETUL KACAR | JAMES KASTING | SCOTT KING | CORBIN KLING | ERIKA KOHLER | SEBASTIAAN KRIJT | CECILIA LEUNG | TIM LICHTENBERG | ROSALY LOPES | MERCEDES LÓPEZ-MORALES | THERESA LUEFTINGER | JACOB LUSTIG-YAEGER | TIM LYONS | CHUHONG MAI | LITON MAJUMDAR | FRANCK MARCHIS | JOSHUA MARTIN | NATHAN MAYNE | FRANCIS MCCUBBIN | KATHLEEN MCINTYRE | VICTORIA MEADOWS | S. RACHEL MELROSE | MOHIT MELWANI-DASWANI | CAMERIAN MILLSAPS | DANTE MINNITI | SARAH MORAN | ANDREAS MORLOK | MIKI NAKAJIMA | ALEXANDRA NAVROTSKY | MARC NEVEU | ATHANASIA NIKOLAOU | LENA NOACK | JESSICA NOVIELLO | JOSEPH O'ROURKE | KAVEH PAHLEVAN | JANI RADEBAUGH | SETH REDFIELD | AMY RICHES | EDGARD RIVERA-VALENTÍN | AKI ROBERGE | MARC ROVIRA-NAVARRO | SARAH RUGHEIMER | ALICIA RUTLEDGE | FRANCESCA SCIPIONI | JOHNNY SEALES | MELISSA SEDLER | DENIS SERGEEV | VINAYAK SHASTRI | HANNAH SHELTON | SANG-HEON DAN SHIM | EVERETT SHOCK | STEVEN SHOLES | RAMDAYAL SINGH | EVAN SNEED | KRISTA SODERLUND | LINDA SPILKER | SIMON STÄHLER | KEVIN STEVENSON | MARSHALL STYCZINSKI | DAVID TOVAR | KEVIN TRINH | PILAR VERGELI | AUSTIN WARE | JUNE WICKS | DAVID WILLIAMS | DAVID WRIGHT | PATRICK YOUNG | QIAN YUAN | MAHEENUZ ZAMAN

## Key Findings

- The search for exoplanetary life must extend beyond a straightforward assessment of habitability as a binary phenomenon, tied to the presence of liquid water, to a systematic understanding of the complex geological processes reflected in an exoplanet's atmosphere—or we risk reporting false positive and false negative detections of exoplanetary life
- We should nurture the nascent discipline of "exogeoscience" to fully integrate astronomers, astrophysicists, geoscientists, oceanographers, atmospheric chemists, and biologists
- Increased funding for interdisciplinary research programs, supporting existing and fostering future multidisciplinary research nodes, and developing research incubators is key to transforming true exogeoscience from an aspiration into reality

## 1. Exoplanet Habitability

The field of exoplanetary science is booming. To date, we have confirmed more than 4,100 extrasolar planets, with that number set to quickly rise thanks to the Transiting Exoplanet Survey Satellite mission. An ever-growing fraction of these exoplanets is rocky. Yet these rocky bodies reveal a radical diversity of types with no counterparts in the Solar System, from Earth-mass worlds with much more water than Earth, such as some in the TRAPPIST-1 system (Grimm *et al.*, 2018; Unterborn *et al.*, 2018) to 55 Cancri e—a super-Earth with a surface temperature greater than the melting point of most rocks (Demory *et al.*, 2016). As the number of known rocky exoplanets grows, exoplanetary science is diligently working to better understand the frequency, composition, and, increasingly, the *geological* nature of these planets—including their potential to be habitable over geological (>1 Gyr) timescales.

A major challenge to that better understanding, however, is the limited available observations for individual exoplanets. In the optimum case, we are at present capable of measuring only an exoplanet's mass, radius, orbital parameters, temperature/phase curves (in rare cases), and the composition of any atmosphere present. Some surface spatial patterning may be observable in the near future (Fujii *et al.*, 2018), and the surface of at least one world, LHS 3844b, has been glimpsed (Kreidberg *et al.*, 2019). The gulf between planetary and exoplanetary science is narrowing, but it is still a gulf.

Happily, in the coming decades several new instruments capable of observing exoplanet atmospheres and surfaces will see first light, including the James Webb Space Telescope (JWST), the Extremely Large Telescope, the Giant Magellan Telescope, and the Atmospheric Remote-sensing Infrared Exoplanet Large-survey. Measuring the composition of a planet's atmosphere will thus grow more routine, coming to be one of *the* primary observables for understanding the nature of rocky planets. These secondary atmospheres—generated primarily by volcanism and degassing, either from partial melting of the interior or during an early magma ocean phase—offer us a probe of a planet's prospective habitability, including direct searches for biosignatures (e.g., Meadows and Seager, 2010; Kiang *et al.*, 2018; Schwieterman *et al.*, 2018; Grenfell, 2019). But interpreting these measurements in terms of detecting life is challenged not only by the presence of clouds, aerosols, and ash plumes, but by the nature and history of the atmosphere itself—and particularly the role of geology in shaping the abiotic background from which we must distinguish prospective biosignatures.

### 1.1. Habitability Versus Detectability

Planets orbiting within the habitable zone of their host star are prime observational targets in our search for exoplanetary life. But are they the only places to search for this life? Life on Earth is not enabled solely because of its stable orbit at 1 AU around a G-type star. Rather, the geological characteristics of the planet play a central role in its habitability, chiefly by facilitating a relatively stable climate over the long term through C, N, and $H_2O$ cycles that exchange material from the interior to the surface and back again via volcanism and subduction (e.g., Kasting and Catling, 2003). Other geological aspects are important, too, including the cycling of nutrients into the oceans as a function of the fraction of exposed land and degree of orogeny (Glaser *et al.*, 2020). An intrinsically generated magnetic field may also play a role in protecting the surface from harmful XUV radiation and minimizing atmospheric stripping (Dong *et al.*, 2020), although the extent to which magnetic fields operate on exoplanets and contribute to habitability remains to be determined (Driscoll and Barnes, 2015; Gunnell *et al.*, 2018).

Geological processes themselves can lead to false negative or false positive observations for individual biosignature-relevant atmospheric species (e.g., $O_2$, $CH_4$). Abiotic $CH_4$ can be produced by water–rock interactions (e.g., Etiope, 2017), and abiotic $O_2$ can build up from water photolysis (Wordsworth and Pierrehumbert, 2014) or because of runaway greenhouse effects (Luger and Barnes, 2015), both of which might result in a false positive detection of life on an exoplanet. Similarly, biologically produced $O_2$ could react with gases released from a reduced mantle (e.g., CO, S), to form $CO_2$ and $SO_2$, in turn leading to a false negative detection (Reinhard *et al.*, 2017; Meadows *et al.*, 2018). **The search for exoplanetary life therefore cannot simply encompass a straightforward assessment of habitability in terms of orbital position, water stability, etc., nor can it ignore the complex geological processes reflected in an exoplanet's atmosphere.** Indeed, the abiotic production and modification of secondary atmospheres represents a confluence of astronomical, geological, and geophysical properties and processes taking place perhaps even before a planetary body's formation has ended. The initial composition of the body, inherited from the portion of the protoplanetary disk in which it formed, sets its mineralogy and volatile content (Raymond *et al.*, 2004; Bond *et al.*, 2010; Unterborn and Panero, 2017). During its magma ocean phase, geochemistry and mineral physics set the stage for segregation of the Fe-rich core, which permanently sequesters biocritical elements such as C, P, and O from the mantle as it grows (Stewart and Schmidt, 2007; Li *et al.*, 2016; Fischer *et al.*, 2020).

As the body solidifies, incompatible elements are partitioned into the crust, leaving behind a depleted mantle that will chemically evolve further by melting and volcanic degassing. Of note, the timing of magma ocean crystallization and devolatilization can vary by hundreds of millions of years because of the effect of tidal heating from companion planets or moons (Renaud and Henning, 2018), so the evolutionary history of an exoplanet need not follow too closely that of worlds in the Solar System.

Through atmosphere–surface interactions, volatiles lock into rocks and are transported potentially all the way to the deep mantle via tectonic processes and mantle convection (Hirschmann, 2006; Dasgupta and Hirschmann, 2010). In parallel, the atmosphere undergoes chemical processes that create and obliterate both biotically and abiotically derived molecules, and interactions with the stellar environment potentially destroy biosignature gasses (e.g., Barabash *et al.*, 2007). All of these phenomena have their own time dependence and likely co-evolve with any life present on these worlds, only to be boiled down to a single, instantaneous atmospheric spectral measurement—with which we must deduce the abiotic history of the planet, its potential habitability in a geological context, and whether life is even detectable in the first place above the abiotic background of that atmosphere.

## *1.2. Unravelling Interior–Surface–Atmosphere Interactions*

Studying Earth itself is central to understanding the abiotic background of atmospheric gasses, but only to a point—our atmosphere is indelibly marked by the signature of life. To understand the abiotic context of rocky exoplanets, then, we must look to models that are informed by experiments, theory, models and, where available, observations. But the creation of these coupled interior–surface–atmosphere models clearly cannot be the purview of any one discipline, and cannot only encompass those planets with which we are familiar. Thus, as we move toward quantifying the detectability of life, **we must work to develop a truly integrated discipline of "exogeoscience" that includes input from** **astronomers, astrophysicists, geologists, geochemists, geodynamicists, meteoriticists, mineral physicists, oceanographers, atmospheric scientists, and biologists, taking into account insights from observationalists, experimentalists, theorists, and modelers alike.**

Building holistic, systems-level models requires fostering this interdisciplinary approach with the ultimate goal of distinguishing the abiotic evolution of an exoplanet from a detected potential biosignature (cf. Barnes *et al.*, 2019). Yet, because of the known diversity of exoplanets (e.g., Selsis *et al.*, 2008), the recognition that Earth-size worlds in orbit about other stars may not in fact be anything like Earth (Kane *et al.*, 2014), the possibility of planetary migration, and perhaps even the rarity of truly Earth-like worlds, it is increasingly clear that such models cannot simply be variations on a theme of Earth. We must devise new experiments and models to better understand and characterize the planetary diversity we now know to exist, and explore the broader parameter space of rocky exoplanets to be able to make sense of those *not* yet known to exist.

## 2. Barriers to Success

The creation of such experiments and models represents a major interdisciplinary undertaking where observers, modelers, and experimentalists must understand the application, and limitations, of each other's contributions to this shared endeavor. Currently, the constituent communities needed to fully investigate interior–surface–atmosphere interactions through time are fragmented, even within the individual geoscience and astronomical disciplines. And although the "traditional" Earth, meteoritic, and planetary science communities sporadically interact, as they collectively seek to understand the origins of Earth and other Solar System bodies, each of these communities has their own preferred journals, conferences, workshops, data formats, and even Decadal Surveys or equivalent flagship professional reports. (To wit, the recent "Earth in Time" National Science Foundation Earth Sciences 2020–2030 consensus study report explicitly listed planetary science as "not in its purview" (p. 23).)

### 2.1. Parallel Communities, Worlds Apart

Both geoscience and astronomy are filled with jargon, sometimes having two substantially different meanings for the same word. For example, to one researcher the word "core" denotes the central, iron-rich core of a planetary body, but to another it can mean the entire planet beneath an atmosphere. Similarly, few astronomers and geoscientists use the term "metal" in the same way. And "Earth-like" is often used to describe planets with some property similar to Earth, but for which the surface and atmospheric conditions may be anything but. Indeed, these fields can have very different science priorities, with planetary geoscience seeking to understand aspects such as composition, density, geodynamics, volatile cycling, and surface features, whereas exoplanetary astronomy focuses on characterizing a body's orbital parameters, its formation, and the age and properties of its host star. There has been, to date, insufficient interaction between exoplanetary astronomers and planetary geologists, whose knowledge of surface processes, petrology, and volcanism will be critical to helping link planetary properties to atmospheric composition and so develop exogeoscience as a viable discipline, especially given the anticipated deluge of such data in the coming decades.

Developing complex, planetary-scale models requires considerable infrastructure development, as well. Crucial experiments on atmospheric parameters, melting behavior, thermoelastic mineral properties, etc., call for major investment in laboratory space. These investments are necessarily expensive and labor intensive, often having to accommodate high pressures and temperatures. High-performance computing (and associated expenditure) is just as important. Without such infrastructure, the vital interior–surface–exosphere models we need going forward will simply not be possible.

## 2.2. A Critical Juncture

In the absence of a truly interdisciplinary exoplanetary field, scientists may not be comfortable branching out beyond their established expertise, especially if they feel that interdisciplinary projects are seen as high risk by funding agencies already facing severe selection pressures. Compounding this view is that interdisciplinary projects can also be perceived as having a low or, at best, uncertain impact in the absence of robust observables, especially for exoplanetary research—despite the enormous work, by definition interdisciplinary, required to obtain those observables in the first place.

These concerns are likely amplified for early career researchers and those without the security of a permanent position. Funding sources such as NASA's Habitable Worlds (HW) and Exoplanets Research (XRP) programs have sought to encourage interdisciplinary research by being cross divisional. Obtaining appropriate reviews can be difficult, however, in part because of bias (or at least preference) toward proposals for which the data needed for the validation of project results are already available.

Nevertheless, there is a pressing need to partner planetary geoscientists and exoplanetary researchers together, or we will not be able to fully understand the ever increasing number of rocky extrasolar worlds we discover. Solar System science experienced a similar paradigm shift in the 1960s, when the advent of robotic missions meant that planets were no longer the exclusive purview of astronomers. This shift required Earth-focused geoscientists to think materially about other worlds; in fact, geoscientists were key to the successful training of the Apollo astronauts (King, 1989). Even within planetary science, the focus on "follow the water" in the characterization of Mars' past habitability showed the importance of incorporating Earth-based expertise, such as sedimentary geology, in the planning, acquisition, and analysis of Mars mission data. We in astronomy and geoscience now face a similar discipline-defining moment, as we seek to interpret a single pixel, a wobble in a star's light, or a dip in brightness through the lens of planetary habitability and biosignature detectability. **Without close collaboration between planetary geoscientists and exoplanetary astronomers, we risk the first celebrated detection of alien life through atmospheric spectroscopy actually being the consequence of a geology we do not understand.**

## 3. The Path Forward

The key to fostering ever-closer collaboration between planetary geoscientists and exoplanetary researchers is ensuring that each community can understand the other. Working to increase this understanding—sharing terms, making readily accessible explanations and definitions for techniques and jargon inherent to each group, and communicating the capabilities and limitations of each discipline (cf. Arney *et al*., 2020; Marley *et al*., 2020)—offers the best chance for developing exogeoscience as a discipline. Equally, geoscientists may not be aware of the key measurements of planetary properties astronomy can provide, and vice versa. It is therefore important to establish what each discipline can provide the other, and how, to help researchers from both communities work together more effectively and innovatively to achieve the shared goal of detecting extrasolar life.

## 3.1. What Can the Astronomy Community Provide to Geoscience?

**Astronomers are the key to** *acquiring* **data for planets outside the Solar System.** There are numerous techniques employed to detect exoplanets, although the transit photometry and radial velocity methods are dominant. These two techniques return measurements of planetary size and mass, respectively, which together can be used to determine planetary type (e.g., super-Earth, mini-Neptune, etc.), inform interior structure models, and place broad bounds on possible compositions, oxygen fugacities, surface conditions, etc. Yet the uncertainty in measurements of planetary mass and radius can be substantial. Follow-up measurements, too, can considerably alter our view of a confirmed exoplanet: Kepler-10c was originally reported to have mass and radius values that would classify it as a new type of massive, rocky planet (Dumusque *et al.*, 2014), but subsequent observations revealed it as a lower-mass, volatile-rich planet (Rajpaul *et al.*, 2017).

Astronomy can offer other critical information on exoplanets. For example, the ages of Sun-like stars are estimated with activity diagnostics, gyrochronology, and lithium depletion (Soderblum, 2010), which, by extension, also yields an upper limit for the age of any exoplanets present—with implications for a planet's thermal and geological evolution (e.g., Kite *et al.*, 2009; Foley and Smye, 2018). Observations of extrasolar systems can also return information on orbital dynamics. Such observations bear on the prospect for tidally induced volcanism, and whether planetary migration influenced the final distribution of bodies within a given system of interest, with attendant implications for composition, volatile inventories, etc. (Renaud and Henning, 2018). When acquired, measurements of atmospheric composition (e.g., from transit photometry/spectroscopy or even direct spectroscopy) can bound estimates of a planet's total volatile inventory, with implications for bulk composition, core chemistry, interior processes, and degassing history. And by considering the characteristics of the host star, it may be possible to understand the UV environment to which the planet is subjected, and even its atmospheric loss rates (Bourrier *et al.*, 2018)—helping place constraints on interior degassing models and whether we should expect to find an atmosphere for a given exoplanet (Zahnle and Catling, 2018). **Equipped with measurements such as these, geoscientists can explore planetary boundary conditions beyond those found in the Solar System at present.**

## *3.2. What Can the Geoscience Community Provide to Astronomy?*

**Geoscientists are the key to *interpreting* data for planets outside the Solar System.** For instance, computer modeling and laboratory experiments can simulate the effects of variable bulk composition, oxygen fugacity, etc., on mantle chemistry, core formation, and interior structure (e.g., Boujibar *et al.*, 2020), with important implications for the chemistry of a planetary surface and even its possible volcanic and tectonic characteristics (e.g., Kite *et al.*, 2009). For instance, Boukaré *et al.* (2019) showed how oxygen fugacity played a major role in the thermochemical evolution of Mercury, a highly reduced planet close to its parent star and, perhaps, a high-density exoplanet archetype (Santerne *et al.*, 2018).

Laboratory and numerical studies can also help extend our understanding of Solar System bodies, including Earth, to the possible surface conditions for the range of known planetary masses, bulk compositions, insolation values, etc. Further, the mineral physics community is well positioned to generate data for planetary materials of non-Earth compositions at pressures and temperatures beyond those relevant for Earth and other worlds in the Solar System (Unterborn and Panero, 2019)—a key step in coming to grips with exoplanets as diverse geological worlds in their own right.

Geophysical modeling can help us tackle another major aspect of planetary behavior: the prospect for recycling volatiles (and nutrients) on exoplanets through plate tectonics or other processes capable of moderating and stabilizing planetary temperature over geological timescales. Whether plate tectonics is more (Valencia *et al.*, 2007; van Heck and Tackley, 2011) or less (O'Neill and Lenardic, 2007; Stein *et al.*, 2011) likely on super-Earths or worlds with non-Earth-like compositions remains an open question. However, recent studies suggest that there are more outcomes possible for planetary interior–surface coupling than simply stagnant or episodic lid tectonics, with planets likely transitioning between these states through time (cf. Lenardic, 2018)—even if substantial uncertainties remains (e.g., Seales and Lenardic, 2020). And it is possible that stagnant lid planets themselves can support volatile cycling in a fashion similar to those with plate tectonics by volcanism and crustal foundering (Foley and Smye, 2018). Indeed, stagnant lids may be the most common tectonic style, and basis for volatile recycling, of rocky worlds generally.

Fully characterizing these different tectonic regimes will help address questions regarding what starts, drives, and stops plate tectonics, the timing of such events, and by what means (and how effectively) volatiles and nutrients can be recycled between land, ocean, and interior reservoirs. Given the close link between tectonic regime and volcanism (e.g., Byrne, 2019), a more comprehensive understanding of the tectonic parameter space possible for rocky exoplanets will provide a basis for interpreting evidence of extrasolar volcanism (e.g., Kaltenegger *et al.*, 2010; Tamburo *et al.*, 2018), and inferring the volcanotectonic character of worlds for which volcanism is suspected (Kreidberg *et al.*, 2019). **Now is the time for geoscience to build a framework for interpreting new exoplanetary observations, and for generating hypotheses to be tested by ever more capable instrumentation.**

## 4. The Next Steps

The drive for interdisciplinarity in exoplanetary science must come from the geoscience and astronomy communities, but can only be effectively realized by NASA's Science Mission Directorate and supported by professional societies, e.g., the American Geophysical Union, the American Astronomical Society, the Geological Society of America, etc. Similar efforts for international meetings, such as the Europlanet Science Congress, would also help enormously. But there is ample opportunity for grass-roots support of exogeoscience: we can advocate for increased interdisciplinary funding of the XRP, HW, Exobiology, etc. programs, interdisciplinary research nodes, e.g., the Nexus for Exoplanet System Science (NExSS) program, and multidisciplinary projects such as NASA's Interdisciplinary Consortia for Astrobiology Research. The successful NSF-supported Cooperative Institute for Dynamic Earth Research (CIDER) research incubator and summer school offers a model for exogeoscience-focused projects.

We can work to increase the scientific diversity of our conversations by introducing Earth scientists, biologists, ecologists, etc., to the problems and promises of exogeoscience, and by partnering with Earth-focused society and union divisions and sections. Several such efforts—the Comparative Climatology of Terrestrial Planets meetings in 2012, 2015, and 2018, the NExSS "Habitable Worlds" meeting in 2017, the AGU–AAS–Kavli Foundation "Frontiers in Exoplanets" session at the 2019 AGU Fall Meeting and the 2020 AAS Winter Meeting, and the "Exoplanets in Our Backyard" meeting in February 2020 (Arney *et al.*, 2020)—are exciting steps towards this discourse. Exogeoscience will further benefit from tightly focused workshops to build a systems-level understanding of planetary interior–surface–atmosphere interactions, featuring organizing committees diverse both scientifically and in actuality.

Geoscientists and astronomers can write papers together. Such partnerships will enhance the reach and utility of exogeoscience studies, with co-authors helping each other to draw reasonable inferences from the limited observables to hand and make testable predictions. These efforts will not only grow an interdisciplinary pool of reviewers for both manuscripts and proposals, but will help to reduce jargon and ultimately aid these disparate groups in better communicating with one another. Equally, we can encourage journals to support interdisciplinary papers through, for example, recruiting editors and reviewers who understand interdisciplinary work, and by hosting special issues and coordinated papers in affiliated journals—for example, the "Exoplanets: The Nexus of Geoscience and Astronomy" special issue in the *Journal of Geophysical Research: Planets*. And promoting efforts by the geoscience community to archive journal articles with online repositories (including, but not limited to, arXiv.org and essoar.org) will help increase the reach of geoscience research to other disciplines.

We face a choice: to continue exoplanetary geoscience and astronomical research as two separate disciplines, or to build toward a shared understanding of our universe by working together. **We are on the cusp of being able to detect potential biosignatures on extrasolar planets, and answering one of the most profound questions of all: Are We Alone?**

Will you join us?

## 5. References

Arney, G. N. et al. (2020) Exoplanets in our Backyard: A report from an interdisciplinary community workshop and a call to combined action. *2023–2032 Decadal Survey White Paper.*

Barabash, S. et al. (2007) The loss of ions from Venus through the plasma wake. *Nature*, 450, 650–653.

Barnes, R. et al. (2019) Geoscience and the search for life beyond the Solar System. *Decadal Survey on Astronomy and Astrophysics 2020 White Paper.*

Bond, J. C., O'Brien, D. P. & Lauretta, D. S. (2010) *Astrophysical Journal*, 715:1050.

Boujibar, A., Driscoll, P. & Fei, Y. (2020) Super-Earth internal structures and initial thermal states. *Journal of Geophysical Research: Planets*, 125, e2019JE006124.

Boukaré, C-E., Parman, S. W., Parmentier, E. M. & Anzures, B. A. (2019) Production and preservation of sulfide layering in Mercury's mantle. *Journal of Geophysical Research: Planets*, 121.

Bourrier, V. et al. (2018) Hubble PanCET: an extended upper atmosphere of neutral hydrogen around GJ 3470b. *Astronomy & Astrophysics*, 620, A147.

Byrne, P. K. (2019) A comparison of inner Solar System volcanism. *Nature Astronomy*, 4, 321–327.

Dasgupta, R. & Hirschmann, M. M. (2010) *Earth and Planetary Science Letters*, 298, 1–13.

Demory, B.-O. et al. (2016) A map of the large day–night temperature gradient of a super-Earth exoplanet. *Nature*, 532, 207–209.

Dong, C., Jin, M. & Lingam, M. (2020) Atmospheric Escape From TOI-700 d: Venus versus Earth Analogs. *Astrophysical Journal Letters*, 896:L24.

Driscoll, P. E. & Barnes, R. (2015) Tidal heating of Earth-like exoplanets around M stars: Thermal, magnetic, and orbital evolutions. *Astrobiology*, 15, 9.

Dumusque, X. et al. (2014) The Kepler-10 planetary system revisited by HARPS-N: A hot rocky world and a solid Neptune-mass planet. *Astrophysical Journal*, 789:154.

Etiop, G. (2017) Abiotic methane in continental serpentinization sites: An overview. *Procedia Earth and Planetary Science*, 17, 9–12.

Fischer, R. A. et al. (2020) The Carbon content of Earth and its core. *Proceedings of the National Academy of Sciences*, 117, 8743–8749.

Foley, B. J. & Smye, A. J. (2018) Carbon cycling and habitability of Earth-sized stagnant lid planets. *Astrobiology*, 18, 7.

Fujii, Y. et al. (2018) Exoplanet biosignatures: Observational prospects. *Astrobiology*, 18, 739–778.

Glaser, D. M. et al. (2020) Detectability of life using oxygen on pelagic planets and water worlds. *Astrophysical Journal*, 893:163.

Grenfell, J. L. (2019) Exoplanetary Biosignatures for Astrobiology. In: Cavalazzi, B. and Westall, F. (eds.) *Biosignatures for Astrobiology*. Advances in Astrobiology and Biogeophysics, Springer, Cham.

Grimm, S. L. et al. (2018) The nature of the TRAPPIST-1 exoplanets. *Astronomy & Astrophysics*, 613, A68.

Gunnell, H. et al. (2018) Why an intrinsic magnetic field does not protect a planet against atmospheric escape. Astronomy and *Astrophysics*, 614, L3.

Hirschmann, M. M. (2006) Water, melting, and the deep Earth H$_2$O cycle. *Annual Review of Earth and Planetary Sciences*, 34, 629–653.

Kaltenegger, L. et al. (2010) Deciphering spectral fingerprints of habitable exoplanets. *Astrobiology*, 10, 1.

Kane, S. R., Kopparapu, R. K. & Domagal-Goldman, S. D. (2014) On the frequency of potential Venus analogs from Kepler data. *Astrophysical Journal Letters*, 794:L5.

Kasting, J. F. & Catling, D. (2003) Evolution of a habitable planet. *Annual Review of Astronomy and Astrophysics*, 41, 429–463.

Kiang, N. Y. et al. (2018) Exoplanet biosignatures: At the dawn of a new era of planetary observations. *Astrobiology*, 18, 619–629.

King, E. A. (1989) Moon Trip: A Personal Account of the Apollo Program and Its Science. Houston, TX: University of Houston, pp. 149.

Kite, E. S., Manga, M. & Gaidos, E. (2009) Geodynamics and rate of volcanism on massive Earth-like planets. *Astrophysical Journal*, 700, 1732–1749.

Kreidberg, L. et al. (2019) Absence of a thick atmosphere on the terrestrial exoplanet LHS 3844b. *Nature*, 573, 87–90, doi:10.1038/s41586-019-1497-4.

Lenardic, A. (2018) The diversity of tectonic modes and thoughts about transitions between them. *Philosophical Transactions of the Royal Society A*, 376, 20170416.

Li, Y., Dasgupta, R., Tsuno, K., Monteleone, B. & Shimizu, N. (2016) Carbon and sulfur budget of the silicate Earth explained by accretion of differentiated planetary embryos. *Nature Geoscience*, 9, 781–785.

Luger, R. and Barnes, R. (2015) Extreme water loss and abiotic O$_2$ buildup on planets throughout the habitable zones of M dwarfs. *Astrobiology*, 15, 119–143.

Marley, M. et al. (2020) Enabling effective exoplanet/planetary collaborative science. *2023–2032 Decadal Survey White Paper.*

Meadows, V. & Seager, S. (2010) Terrestrial planet atmospheres and biosignatures. In *Exoplanets*, Seager S. (ed.), University of Arizona Press, Tucson, pp. 441–470.

Meadows, V. S. et al. (2018) Exoplanet biosignatures: Understanding oxygen as a biosignature in the context of its environment. *Astrobiology*, 18, 630–662.

O'Neill, C. & Lenardic, A. (2007) Geological consequences of super-sized Earths. *Geophysical Research Letters*, 34, L19204.

Rajpaul, V., Buchhave, L. A. & Aigrain, S. (2017) Pinning down the mass of Kepler-10c: the importance of sampling and model comparison. *Monthly Notices of the Royal Astronomical Society: Letters*, 471, L125–L130.

Raymond, S. N., Quinn, T. & Lunine, J. (2004) Making other Earths: Dynamical simulations of terrestrial planet formation and water delivery. *Icarus*, 168, 1–17.

Reinhard, C. T. et al. (2017) False negatives for remote life detection on ocean-bearing planets: Lessons from the early Earth. *Astrobiology*, 17, 287–297.

Renaud, J. P. & Henning, W. G. (2018) Increased tidal dissipation using advanced rheological models: Implications for Io and tidally active exoplanets. *Astrophysical Journal*, 857:98.

Santerne, A. et al. (2018) An Earth-sized exoplanet with a Mercury-like composition. *Nature Astronomy*, 2, 393–400.

Schwieterman, E. W. et al. (2018) Exoplanet biosignatures: A review of remotely detectable signs of life. *Astrobiology*, 18, 663–708.

Selsis, F., Kaltenegger, L. & Paillet, J. (2008) Terrestrial exoplanets: diversity, habitability and characterization. *Physica Scripta*, T130, 014032.

Soderblum, D. R. (2010) The ages of stars. *Annual Review of Astronomy and Astrophysics*, 48, 581–629.

Stein, C., Finnenkötter, A., Lowman, J. P. & Hansen, U. (2011) The pressure-weakening effect in super-Earths: Consequences of a decrease in lower mantle viscosity on surface dynamics. *Geophysical Research Letters*, 38, L21201.

Stewart, A. J. & Schmidt, M. W. (2007) Sulfur and phosphorus in the Earth's core: The Fe–P–S system at 23 GPa. *Geophysical Research Letters*, 34, L13201.

Tamburo, P., Mandell, A., Deming, D. & Garhart, E. (2018) Confirming variability in the secondary eclipse depth of the super-Earth 55 Cancri e. *Astronomical Journal*, 155, 5.

Unterborn, C. T. & Panero, W. R. (2017) The effects of Mg/Si on the exoplanetary refractory oxygen budget. *Astrophysical Journal*, 845, 1.

Unterborn, C. T. & Panero, W. R. (2019) The pressure and temperature limits of likely rocky exoplanets. *Journal of Geophysical Research: Planets*, 124, 1704–1716.

Unterborn, C. T., Desch, S. J., Hinkel, N. R. & Lorenzo, A. (2018) Inward migration of the TRAPPIST-1 planets as inferred from their water-rich compositions. *Nature Astronomy*, 2, 297–302.

Valencia, D., O'Connell, R. J. & Sasselov, D. D. (2007) Inevitability of plate tectonics on super-Earths. *Astrophysical Journal*, 670, L45–L48.

van Heck, H. J. & Tackley, P. J. (2011) Plate tectonics on super-Earths: Equally or more likely than on Earth. *Earth and Planetary Science Letters*, 310, 252–261.

Wordsworth, R. & Pierrehumbert, R. (2014) Abiotic oxygen-dominated atmospheres on terrestrial habitable zone planets. *Astrophysical Journal Letters*, 785, L20.

Zahnle, K. J. & Catling, D. C. (2018) The Cosmic Shoreline: The evidence that escape determines which planets have atmospheres, and what this may mean for Proxima Centauri B. *Astrophysical Journal*, 843:122.